# An echo of the low-temperature phase competition and 1/8 dip in room-temperature properties of cuprate HTSCs

A.V. Fetisov[1]

**Abstract** Cuprate high-temperature superconductors (HTSCs) exhibit a characteristic dome-shaped dependence of the superconducting transition temperature on the charge carrier concentration, $T_c(p)$, featuring a maximum at optimal doping $p \approx 0.16$. Near the $p = 1/8$ composition, a sharp suppression ("dip") of $T_c$ occurs – conventionally referred to as the "1/8 anomaly" – which is widely attributed to charge and spin ordering in the $CuO_2$ planes. Here, we investigate the room temperature hydration of $La_{2-x}Ba_xCuO_4$ and $YBa_2Cu_3O_{6+\delta}$ under a high-frequency magnetic field and report anomalous weight changes during the initial stage of the process. For both compounds, the doping dependence of these weight fluctuations closely mirrors their respective $T_c(p)$ profiles, including the signature "1/8 anomaly". The prominent manifestation of these characteristically low-temperature electronic features at room temperature provides critical insights into the stability of high-energy electronic correlations. These findings may offer a novel foundation for the development of theoretical models of cuprate superconductivity.

## Introduction

One of the most prominent features of cuprate high-temperature superconductors (HTSC) is the dome-shaped dependence of the critical transition temperature $T_c$ on the hole carrier concentration $p$, which typically peaks at the optimal doping level of $p \approx 0.16$ [1, 2]. However, in several systems, such as $La_{2-x}Ba_xCuO_4$ (LBCO), a sharp anomalous suppression ("dip") of $T_c$ is observed within a narrow doping range around $p = 1/8$ [3–5]. In other systems, including $YBa_2Cu_3O_{6+\delta}$ (YBCO), a similar suppression of $T_c$ manifests, albeit being less pronounced and smeared over a wider interval centered near $p = 1/8$ [6–8]. This phenomenon, widely known as the "1/8 anomaly", signifies a fundamental competition between superconductivity and alternative forms of electronic ordering. In cuprate HTSCs, this anomaly is driven by the presence of fluctuating charge density waves (CDWs) in the $CuO_2$ structural planes (as documented for YBCO [9–11]) or alternating charge and magnetic "stripes" (as seen in LBCO [12]). At the specific 1/8 doping, their wave vectors resonate with the periodicity of the crystal lattice, locking into a static state and, in some instances, stabilizing a three-dimensional long-range order. As these ordered phases strengthen, they disrupt the phase coherence of the superconducting order parameter (SC), thereby suppressing $T_c$, occasionally down to zero [2, 13].

[1] Institute of Metallurgy of the Ural Branch of the Russian Academy of Sciences, Ekaterinburg, Russian Federation, fetisovav@mail.ru

On the phase diagram of cuprate HTSCs, a quantum critical point (QCP) is located at $p^* \cong 0.19$, $T = 0$ [14]. At this point, quantum magnetic fluctuations reach their maximum and are believed to act in tandem with phonons to mediate hole pairing in the $CuO_2$ planes, leading to the highest $T_c$ values near $p^*$ (at optimal doping $p \approx 0.16$). With increasing temperature, the doping range governed by these quantum fluctuations expands, defining the "strange metal" phase, where spin fluctuations persist well above room temperature (RT) [15, 16]. At $T$ = RT, this phase manifests as a broad, homogeneous region centered at $p \sim 0.19$, encompassing both the $p \sim 0.16$ and $p \sim 1/8$ regions without exhibiting any singular features at these points [2, 15]. Nevertheless, over the last decade, numerous studies have identified dynamic charge density fluctuations within the strange metal phase (see, e.g., [16, 17] and references therein), although long-range ordered electronic or magnetic phases are typically absent at these elevated temperatures [2, 15, 16].

In the present study, we demonstrate that specific doping intervals near $p$ = 1/8 and $p$ = 0.16 – closely matching the characteristic features of the $T_c(x/\delta)$ dependence – manifest at temperatures significantly higher than previously assumed. In our experiments on the cuprate HTSCs $La_{2-x}Ba_xCuO_4$ and $YBa_2Cu_3O_{6+\delta}$, these features emerge as an exotic weight loss in the samples during room-temperature hydration and under the influence of a high-frequency magnetic field. Notably, on the phase diagram, this temperature is far above both $T_c$ and the onset temperature of CDW formation (100 K < $T_{CDW}$ < 200 K). The objective of this work is to highlight the specific behaviors of HTSCs near 1/8 and 0.16 doping points, suggesting they stem from electronic or structural correlations that are inherently more robust than CDW and superconductivity alone.

This paper provides systematic experimental data on the sharp weight loss of gas-tight reactors containing powdered HTSCs and a crystal hydrate that stabilizes the humid atmosphere. We show that the magnitude of this weight loss, as a function of the HTSC composition, strictly replicates the topology of the well-known dome-shaped $T_c(x/\delta)$ curves, including the deep "dip" near $p$ = 1/8. This investigation extends the author's previous studies [18, 19] into the nature of high-temperature anomalous behaviors in cuprates.

## Experimental Details

The experimental procedure for the YBCO compound is described in detail in our previous works [18, 19]. In this study, we supplemented those results with weight loss data for two new oxygen compositions ($\delta$ = 0.27 and 0.90).

The cuprate LBCO (with compositions $x$ = 0.05, 0.10, 0.1125, 0.125, 0.1375, 0.15, 0.20, and 0.25) was synthesized from $La_2O_3$, CuO, and barium carbonate using a conventional solid-state reaction method. The initial calcination of stoichiometric mixtures was performed at 900 °C for 20 hours. After regrinding, the mix-tures were fired at 1100 °C for 20 hours, and then at 1080 °C for 45 hours with intermediate grinding. The synthesized material was subsequently oxidized in an oxygen atmosphere at 510 °C for 3 hours. All following procedures – including sample preparation, loading into a gas-tight reactor, treatment in an alternating magnetic field, and weighing – were carried out in strict accordance with the protocols previously established for YBCO samples, which are described in detail elsewhere [19].

Figure 1 shows the X-ray diffraction (XRD) patterns obtained for the LBCO samples with various barium concentrations $x$. The measurements were carried out using a Shimadzu XRD-7000 diffractometer with $CuK_\alpha$ radiation over a $2\theta$ range of 10–80° with a step size of 0.02° and a counting time of 2 s per step. Phase analysis revealed that all studied lanthanum cuprate compositions are single-phase. No structural anomalies or secondary phases were observed for the $x$ = 0.125 and 0.1375 samples, which exhibited unusual behavior in subsequent measurements.

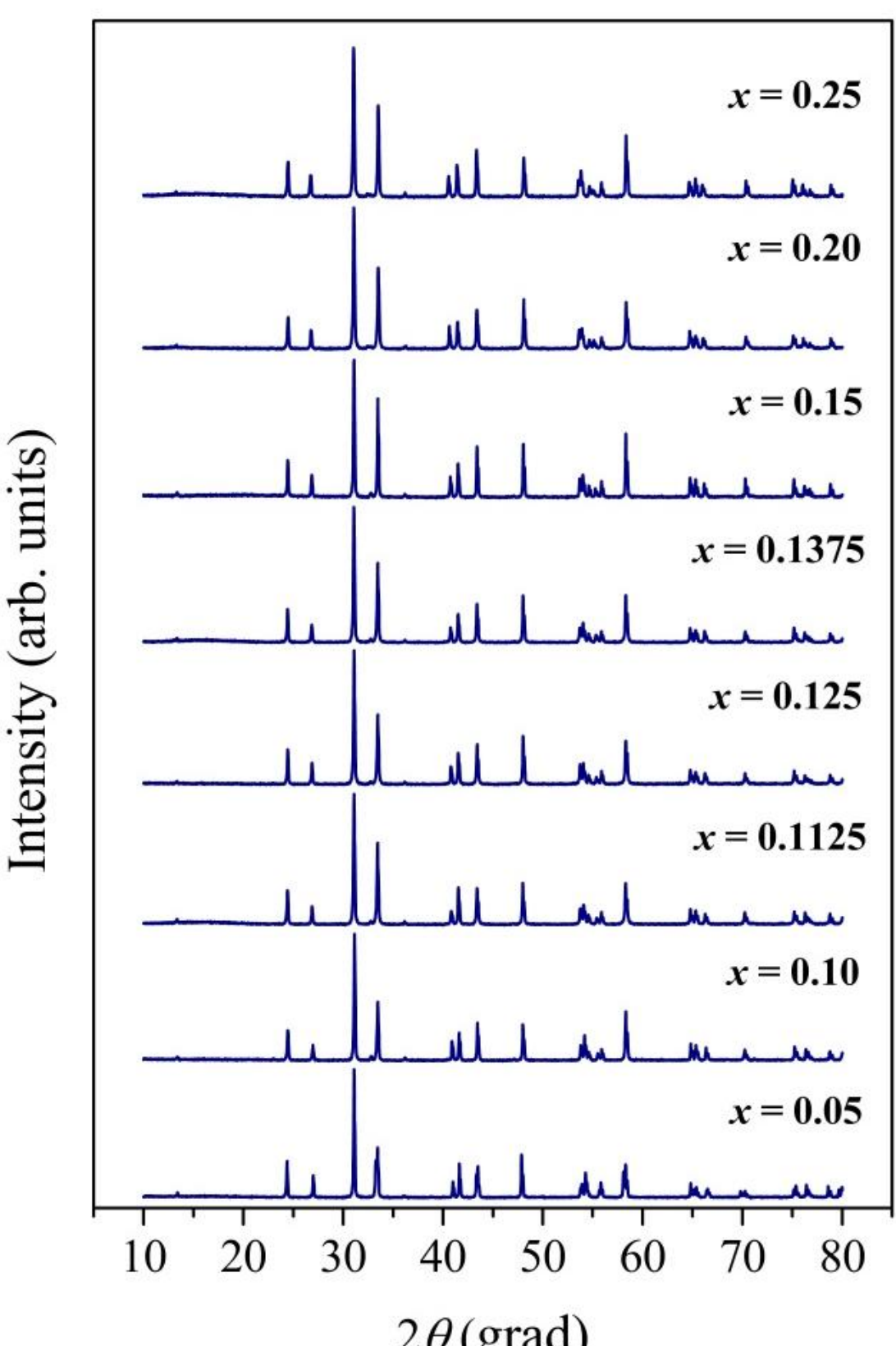

**Fig. 1** XRD characterization of the $La_{2-x}Ba_xCuO_4$ samples.

## Experimental Results

Figure 2 presents the measured weight changes (ΔW) of the reactors containing the LBCO samples. A nearly dome-shaped dependence of these changes on the barium substitution degree ($x$) is clearly observed. This behavior – plotted as the magnitude of the “drop-effect” versus $x$ – is further illustrated in Fig. 3. Notably, a significant dip in the ΔW values occurs at $x$ = 0.125 and 0.1375. Given that the free charge carrier (hole) concentration in LBCO equals the substitution level ($p = x$), this dip takes place at a hole concentration close to 1/8, whereas the maximum

ΔW is reached near $p = 0.16$. A comparison with literature data for $T_c$ reveals that a similar dip and maximum in the $T_c(x)$ dependence are observed at these exact same concentrations, respectively (Fig. 3).

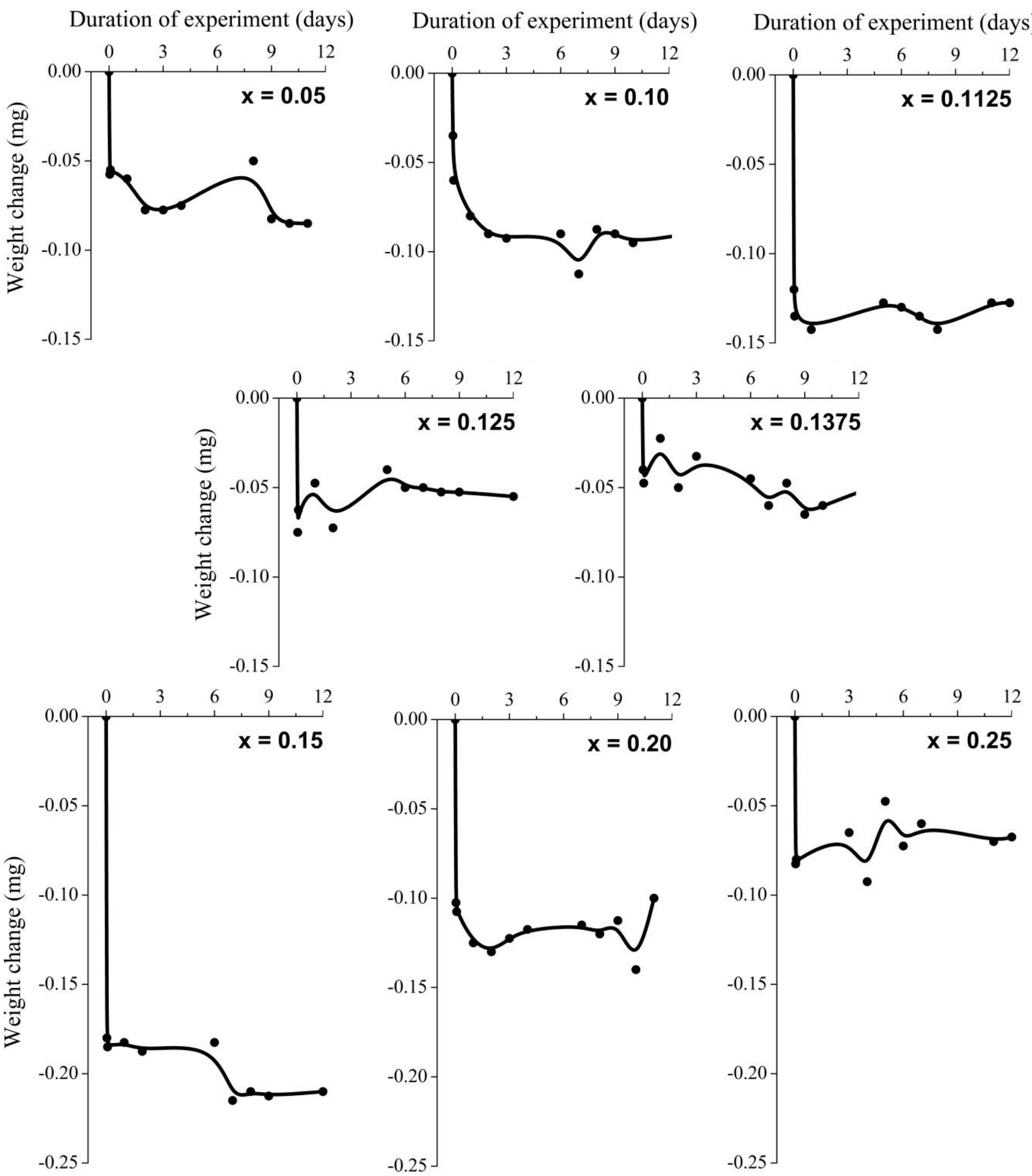

**Fig. 2** Weight loss of reactors with $La_{2-x}Ba_xCuO_4$ samples as a function of time and the barium substitution degree $x$.

In our recent work [19], we reported the ΔW dependencies obtained for YBCO samples with various oxygen indices (δ). The generalized dependence for these data, along with the results for two samples with $\delta = 0.27$ and 0.90 measured subsequently, is presented in Fig. 4. This figure also shows the $T_c(\delta)$ dependences obtained for YBCO by various authors. Although the two curves do not coincide exactly, they clearly share common features: the maximum values of both quantities fall within the δ = 0.90–0.95 interval, while the center of the dip is located at δ = 0.7–0.75. According to the established relationship between δ and $p$ [1], these regions correspond to hole concentrations close to $p$ = 0.16 and 1/8 respectively.

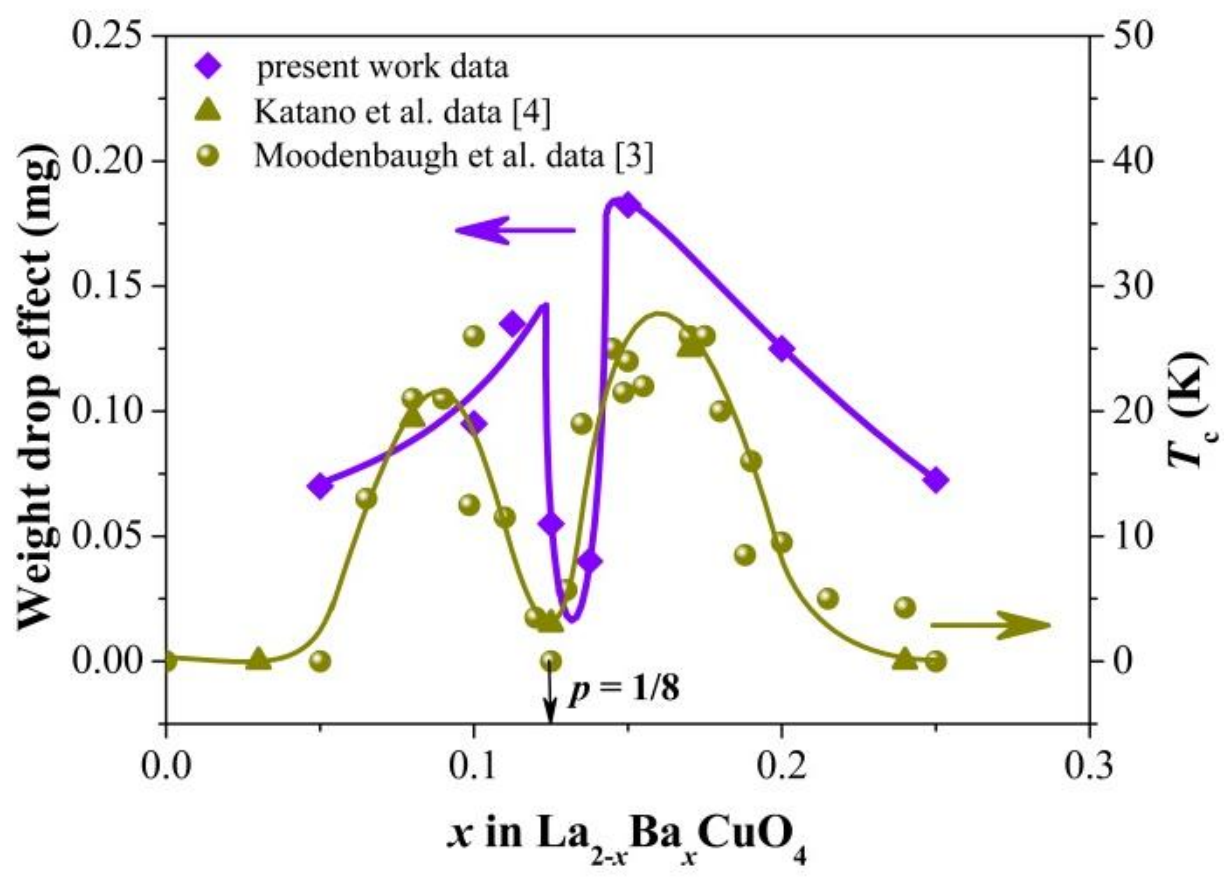

**Fig. 3** Comparison of the drop-effect magnitude and $T_c$ as functions of $x$ for the LBCO compound. The error in the drop-effect values is typically ±0.015 mg.

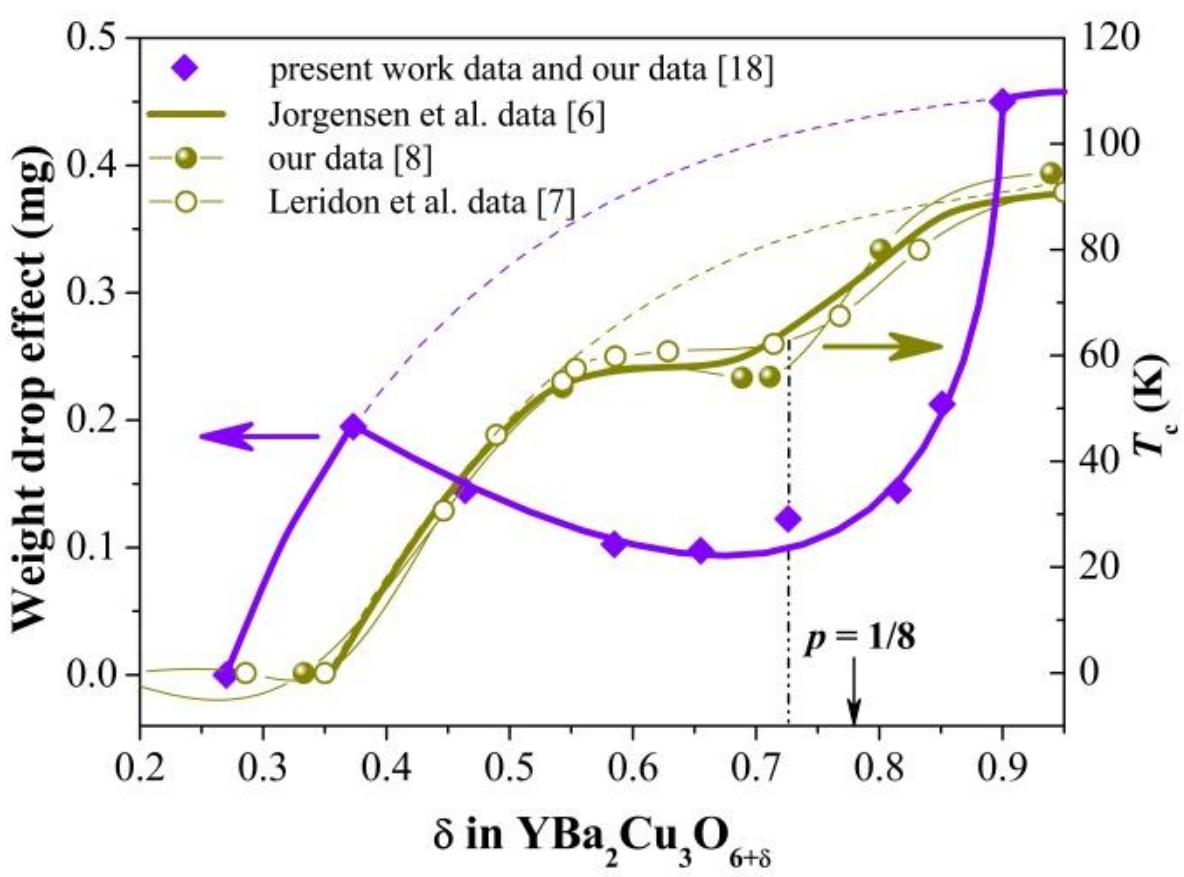

**Fig. 4** Comparison of the drop-effect magnitude and $T_c$ as functions of $x$ for the YBCO compound. The error in the drop-effect values is typically ±0.015 mg.

# Discussion

The primary experimental finding of this work is the anomalous, systematic weight loss ΔW observed in gas-tight reactors during the initial stage of room-temperature hydration of HTSC powders. Crucially, the magnitude of ΔW exhibits a highly non-monotonic dependence on the doping level for both the LBCO and YBCO systems. When plotted against the hole concentration $p$, the weight loss curves closely replicate the characteristic topology of the superconducting transition temperature dependence, $T_c(p)$. Specifically, the maximum weight loss occurs near the optimal doping level $p$ = 0.16, mimicking the dome-shaped profile of the $T_c(x/\delta)$ dependences. Most remarkably, a pronounced suppression of this drop-effect is observed in immediate proximity to the p = 1/8 (0.125) composition. In LBCO, this dip is sharp and deep, whereas in YBCO, it appears as a broad, statistically significant local minimum.

This structural mimicry at $T$ = 295 K is highly unconventional. Neither charge density waves (CDWs) nor superconducting phases exist at room temperature: the macroscopic superconducting order parameter is entirely absent, and static CDWs or stripe phases – typically held responsible for the low-temperature 1/8 anomaly – are thermally melted into a disordered state. Meanwhile, recent studies have discovered that an extensive region of short-range charge modulations persists well above the CDW onset temperature $100 < T_{CDW} < 200$ K [16, 17, 20].

Furthermore, anomalous weight changes in cuprate HTSCs were historically reported by Podkletnov at temperatures $T < T_c$ [21, 22]. While that effect vanished or decreased below the detection limit above $T_c$ under challenging weighing conditions, it clearly associated the drop-effect with the HTSC electronic state. Our experiments demonstrate that the maximum weight loss values at room temperature can reach 0.03% of the sample mass. Therefore, the superconducting phase region appears to possess a high-temperature “echo”. Indeed, quantum magnetic fluctuations, which serve as the pairing mechanism for charge carriers, persist within the strange metal region up to the crystal destruction temperature [15, 16]. It has been hypothesized that fluctuations of the superconducting order parameter may survive within the strange metal phase even at room temperature [23–26].

Our results strongly imply that these high-energy electronic correlations remain highly sensitive to the exact matching between the electronic filling factor and the underlying crystal lattice at the $p \approx 1/8$ and $p \approx 0.16$ singular points. The fact that the 1/8 anomaly manifests so clearly at room-temperature suggests that the underlying phase competition in cuprates possesses a much higher energy scale than previously deduced from low-temperature transport measurements.

## Conclusion

In summary, we have investigated the sharp weight loss (drop-effect) of $La_{2-x}Ba_xCuO_4$ and $YBa_2Cu_3O_{6+\delta}$ samples during room-temperature hydration under an alternating magnetic field. The magnitude of this drop-effect exhibits a non-monotonic dependences on the chemical substitution degree, which almost explicitly replicates the characteristic $T_c(x/\delta)$ curves. The persistence of these electronic signatures at room temperature suggests the existence of robust, high-energy electronic correlations well above the superconducting transition, providing valuable experimental insights for emerging HTSC theories.

The identification of high-temperature charge modulations has required state-of-the-art experimental techniques and substantial effort. It is anticipated that further investigations will also lead to the explicit detection of charge-pairing fluctuations.

### Acknowledgments

The author expresses gratitude to colleagues O.M. Fedorova and T.I. Filinkova for providing X-ray diffraction data of the LBCO and YBCO samples.

## References

1. J. L. Tallon, C. Bernhard, H. Shaked, R. L. Hitterman, J. D. Jorgensen, Generic superconducting phase behavior in high-$T_c$ cuprates: $T_c$ variation with hole concentration in $YBa_2Cu_3O_{7-\delta}$, *Phys. Rev. B* **51,** 12911–12914 (1995). **DOI: https://doi.org/10.1103/PhysRevB.51.12911.**

2. B. Keimer, S. A. Kivelson, M. R. Norman, S. Uchida, J. Zaanen, From quantum matter to high-temperature superconductivity in copper oxides, *Nature* **518**, 179–186 (2015). **DOI: 10.1038/nature14165.**
3. A. R. Moodenbaugh, Y. Xu, M. Suenaga, T. J. Folkerts, R. N. Shelton, Superconuducting properties of $La_{2-x}Ba_xCuO_4$, *Phys. Rev. B* **38,** 4596–4600 (1988). **DOI: 10.1103/PhysRevB.38.4596.**
4. S. Katano, J. A. Fernandez-Baca, S. Funahashi, N. Môri, Y. Ueda, K. Koga, Crystal structure and superconductivity of $La_{2-x}Ba_xCuO_4$ ($0.03 \le x \le 0.24$), *Phys. C.: Supercond.* **214,** 64–72 (1993). **DOI: 10.1016/0921-4534(93)90108-3.**
5. K. Kumagai, Y. Nakamura, I. Watanabe, Y. Nakamichi, H. Nakajima, Strong dependence of the linear-temperature term of heat capacity on Ba- and Sr-doping in the $La_2CuO_4$ system, *J. Magnetism and Magnetic Materials* **76&77,** 601–603 (1988). **DOI: 10.1016/0304-8853(88)90499-4.**
6. J. D. Jorgensen, B. W. Veal, A. P. Paulikas, L. J. Nowicki, G. W. Crabtree, H. Claus, W. K. Kwok, Structural properties of oxygen-deficient $YBa_2Cu_3O_{7-\delta}$. *Phys. Rev. B* **41,** 1863–1877 (1990), **DOI: 10.1103/PhysRevB.41.1863**
7. B. Leridon, P. Monod, D. Colson, A. Forget, Thermodynamic signature of a phase transition in the pseudogap phase of $YBa_2Cu_3O_x$ high-$T_c$ superconductor. *A Letters Journal Exploring the Frontiers of Physics* **87,** 17011 (pp1–6) (2009), **DOI: 10.1209/0295-5075/87/17011.**
8. A. V. Fetisov, G. A. Kozhina, S. Kh. Estemirova, V. Ya. Mitrofanov, On the room-temperature aging effects in $YBa_2Cu_3O_{6+\delta}$, *Phys. C.: Supercond.* **515**, 54–61 (2015). **DOI: 10.1016/j.physc.2015.05.008.**
9. S. Gerber, H. Jang, H. Nojiri, et al., Three-dimensional charge density wave order in $YBa_2Cu_3O_{6.67}$ at high magnetic fields, *Science* **350**, 949–952 (2015). **DOI: 10.1126/science.aac4594.**
10. R. Comin, A. Damascelli, Resonant X-Ray Scattering Studies of Charge Order in Cuprates, *Annual Review of Condensed Matter Physics* **7,** 369–405 (2016). **DOI: 10.1146/annurev-conmatphys-031115-011401.**
11. M. Akoshima, Y. Koike, On the 60 K plateau of $T_c$ in $Y_{1-x}Ca_xBa_2Cu_3O_{7-\delta}$ in relation to the 1/8 problem, *J. of the Physical Society of Japan* **67,** 3653–3654 (1998). **DOI: 10.1143/jpsj.67.3653.**
12. J. M. Tranquada, B. J. Sternlieb, J. D. Axe, Y. Nakamura, and S. Uchida, Evidence for stripe correlations of spins and holes in copper-oxide superconductors, *Nature* **375**, 561–563 (1995). **DOI: 10.1038/375561a0**; J. M. Tranquada, Stripe correlations of spins and holes in copper oxide superconductors, *Neutron News* **7:1,** 17-20 (1996), **DOI: 10.1080/10448639608217720.**

13. M. Vojta, Lattice-symmetry-breaking phenomena in cuprate superconductors: stripes, nematicity, and density waves, *Advances in Physics* **58**, 699–820 (2009). **DOI: 10.1080/00018730903122242.**
14. C. Proust, L. Taillefer, The Remarkable underlying ground state of cuprate superconductors, *Annu. Rev. Condens. Matter Phys.* **10,** 409–429 (2019). **DOI: 10.1146/annurev-conmatphys-031218-013210.**
15. A. Legros, S. Benhabib, W. Tabis, et al., Universal T-linear resistivity and Planckian dissipation in overdoped cuprates, *Nature Physics* **15**, 142–147 (2019). **DOI: 10.1038/s41567-018-0334-2.**
16. R. Arpaia, S. Caprara, R. Fumagalli, G. De Vecchi, Y. Y. Peng, E. Andersson, D. Betto, G. M. De Luca, N.B. Brookes, F. Lombardi, M. Salluzzo, L. Braicovich, C. Di Castro, M. Grilli, G. Ghiringhelli, Dynamical charge density fluctuation pervading the phase diagram of a Cu-based high-$T_c$ superconductor, *Science* **365,** 906–910 (2019). **DOI: 10.1126/science.aav1315.**
17. C. C. Tam, M. Zhu, M. C. Barthélemy, L. J. Cane, O. J. Lipscombe, S. Agrestini, J. Choi, M. Garcia-Fernandez, Ke-Jin Zhou, S. M. Hayden, Doping evolution of the charge density wave and charge density fluctuations in $La_{2-x}Sr_xCuO_4$, *Phys. Rev. B* **113,** 174506 (2026), **DOI: 10.1103/6hcs-2y4y.**
18. A. V. Fetisov, Weight balance violation during hydration of $YBa_2Cu_3O_{6+\delta}$, *J. Supercond. Nov. Magn.* **34,** 2725–2732 (2021), **DOI: 10.1007/s10948-021-05979-8.**
19. A. V. Fetisov, Strange changes in weight of YBCO occurring during its hydration and under the influence of alternating magnetic field, *J. Supercond. Nov. Magn.* **38,** 204 (2025). **DOI: 10.1007/s10948-025-07043-1.**
20. S. M. Hayden, J. M. Tranquada, Charge correlations in cuprate superconductors, *Annu. Rev. Condens Matter Phys.* **15,** 215–235 (2024). **DOI: 10.1146/annurev-conmat-phys-032922-094430.**

21. E. Podkletnov, R. Nieminen, A possibility of gravitational force shielding by bulk $YBa_2Cu_3O_{7-x}$ superconductor, *Phys. C: Supercond.* **203** 441–444 (1992), **DOI: 10.1016/0921-4534(92)90055-H.**
22. E. Podkletnov, Weak gravitational shielding properties of composite bulk $YBa_2Cu_3O_{7-x}$ superconductor below 70 K under an EM field, MSU-chem-95-cond-mat/9701074 5 Feb 1997.
23. N. E. Hussey, High-temperature superconductivity and strange metallicity: simple observations with (possibly) profound implications, *Phys. C: Supercond.* **614,** 1354362 (2023). **DOI: 10.1016/j.physc.2023.1354362.**
24. V. J. Emery, S. A. Kivelson, Importance of phase fluctuations in superconductors with small superfluid density, *Nature* **374,** 434–437 (1995). **DOI: 10.1038/374434a0.**
25. M. Mitrano, A. A. Husain, S. Vig, A. Kogar, M. S. Rak, S. I. Rubeck, J. Schmalian, B. Uchoa, G. D. Gu, P. Abbamonte, Anomalous density fluctua-

tions in a strange metal, *Communications Physics* **6,** 132 (2018). **DOI: 10.1038/s42005-023-01202-6.**

26. C. Jiang, G. A. Ummarino, M. Baggioli, E. Liarokapis, A. Zaccone, Correlation between optical phonon softening and superconducting $T_c$ in $YBa_2Cu_3O_x$ within *d*-wave Eliashberg theory, *J. Phys. Mater.* **7**, 045002 (2024). **DOI: 10.1088/2515-7639/ad6c7f.**